\documentclass[aps,pra,twocolumn,superscriptaddress]{revtex4-1}
\usepackage{natbib}
\usepackage[breaklinks]{hyperref}

\usepackage{graphicx}  % needed for figures
\usepackage{amsmath}
\usepackage{siunitx}
\begin{document}

\title{Breakdown flash at telecom wavelengths in InGaAs avalanche photodiodes}

\author{Yicheng Shi}
\affiliation{Center for Quantum Technologies, 3 Science Drive 2, Singapore
  117543}
\author{Janet Zheng Jie Lim}
\affiliation{Center for Quantum Technologies, 3 Science Drive 2, Singapore
  117543}
\author{Hou Shun Poh}
\affiliation{Center for Quantum Technologies, 3 Science Drive 2, Singapore
  117543}
\author{Peng Kian Tan}
\affiliation{Center for Quantum Technologies, 3 Science Drive 2, Singapore
  117543}
\author{Peiyu Amelia Tan}
\affiliation{Center for Quantum Technologies, 3 Science Drive 2, Singapore
  117543}
\affiliation{Singapore Telecommunications Limited, 31 Exeter Road Comcentre
  \#15-00, Singapore 239732}
\author{Alexander Ling}
\affiliation{Center for Quantum Technologies, 3 Science Drive 2, Singapore
  117543}
\affiliation{Department of Physics, National University of Singapore, 2
  Science Drive 3, Singapore, 117542}
\author{Christian Kurtsiefer}
\email{phyck@nus.edu.sg}
\affiliation{Center for Quantum Technologies, 3 Science Drive 2, Singapore
  117543}
\affiliation{Department of Physics, National University of Singapore, 2
  Science Drive 3, Singapore, 117542}

%\email{\authormark{*}phyck@nus.edu.sg}
%\date{\today}
%\bibliographystyle{unsrt}
%\bibliography{references/reference}

\begin{abstract}
Quantum key distribution (QKD) at telecom wavelengths ($1260-1625\,\rm{nm}$) has the potential for fast deployment due to existing optical fibre infrastructure and mature telecom technologies. At these wavelengths, indium gallium arsenide (InGaAs) avalanche photodiode (APD) based detectors are the preferred choice for photon detection. Similar to their silicon counterparts used at shorter wavelengths, they exhibit fluorescence from recombination of electron-hole pairs generated in the avalanche breakdown process. This fluorescence may open side channels for attacks on QKD systems. Here, we characterize the breakdown fluorescence from two commercial InGaAs single photon counting modules, and find a spectral distribution between $1000\,\rm{nm}$ and $1600\,\rm{nm}$. We also show that by spectral filtering, this side channel can be efficiently suppressed.
\end{abstract}

\maketitle

\section{Introduction}

Quantum key distribution (QKD) enables two distant parties to share a random
encryption key without being eavesdropped by a malicious third party. Since
the proposal of BB84 protocol by Bennett and Brassard in
1984~\cite{Bennett1984}, years of research effort have been committed to
building more efficient and robust QKD systems. These systems can be
implemented using photons transmitted over free space~\cite{Buttler1998, Marcikic2006, Jouguet2012} or over optical fibres~\cite{Townsend1999, Bourennane1999,Ribordy2000}.

QKD implementations over optical fibres receive growing interest due to their
compatibility with existing telecom fibre
networks~\cite{Hughes2000}, but require detection of single photons at
telecom wavelengths ($1260-1625\,\rm{nm}$). Avalanche photodiodes
(APDs) based on Indium Gallium Arsenide (InGaAs) are the commonly used
detectors for this wavelength range~\cite{Muller1997, Stucki2001,
  Hadfield2009}. Despite their relatively high dark count rate as compared to
their silicon counterparts, InGaAs APDs are able to detect single photons at
telecom wavelengths with quantum efficiency up to $20\%$~\cite{Hadfield2009,
  Zhang2015}.

One aspect of APD may provide a susceptibility to the so-called side
channel attack in QKD implementations: the emission of fluorescence light
during the avalanche breakdown process. This light emission is due to the
recombination of electrons and holes in the APD  junction, and covers a
spectrum ranging from $700\,\rm{nm}$ to $1000\,\rm{nm}$ in silicon based APDs.
A similar florescence has also been
observed in InGaAs APDs~\cite{Marini2016}. This fluorescence light (referred
to as 'breakdown flash') reveals information about the photon
detection process to an eavesdropper~\cite{Kurtsiefer2001}. Depending on the
detection scheme in a quantum communication setting, the eavesdropper may
extract timing and/or polarization information of the detected photons by
observing the breakdown flash leaked back to the optical channel. Thus, a
strategy must be in place to reduce or eliminate this side channel.

In this work, we investigate the breakdown flash from two commercial InGaAs
single-photon counting modules (ID220, ID Quantique). In doing so, we obtain a
lower bound for the breakdown flash probability. We also characterize its
spectral distribution and find that by spectral filtering, the number of
detected 
breakdown flash events can be greatly suppressed. Under the conservative
assumption on a unit detector efficiency, one can place an upper bound to the
information leakage due to the breakdown flash in a QKD scenario.

\section{Detection of breakdown flash}

\begin{figure*}%[!ht]
\centering%\includegraphics[width=0.9\columnwidth]{./plots/setup_collimator/setup_collimator_fibre_length}
\includegraphics[width=1.5\columnwidth]{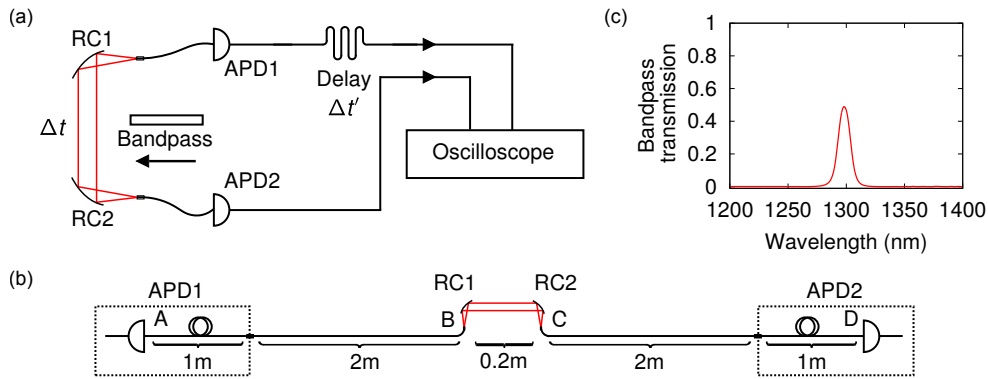}
\caption{(a) Setup for detecting the breakdown flash. The two APDs are
  optically coupled to each other by a pair of reflective collimators (RC1 and
  RC2). It takes $\Delta t \approx 32.5\,\rm{ns}$ for a photon to travel the
  optical distance between APD1 and APD2. (b) Schematics of the lengths of
  fibre patchcords. The output signals from APDs are sent to an oscilloscope
  with an electrical delay $\Delta t' \approx 127\,\rm{ns}$ applied to
  APD1. The oscilloscope triggers on signals received from APD2, and records
  the arrival times of signals from APD1. We record coincidence both events
  where APD1 emits a breakdown flash that is detected by APD2, and the other
  way round. An optical bandpass filter in a another measurement
  to suppress the number of breakdown flash events. The transmission profile
  of the bandpass filter is shown in (c).}
\label{fig:setup_collimator}
\end{figure*}

%The setup for detecting breakdown flash is shown in Fig.~\ref{fig:setup_collimator}. Two commerical single-photon counting modules APD1 and APD2 (ID220 from ID Quantique) are optically coupled to a pair of reflective collimators (RC1 and RC2) looking at each other throught freespace. The optical distance between the two APDs is about $9.6\,\rm{m}$ ($9.5\,\rm{m}$ fibre $+ 0.1\,\rm{m}$ free space) which corresponds to a photon traveling time of $\Delta t \approx 48\,\rm{ns}$.

The devices under test are two InGaAs APD based single-photon counting
modules, APD1 and APD2 (ID220, ID Quantique, with fibre input). We use the
setup shown in Fig.~\ref{fig:setup_collimator} where each counting module acts
as both source and detector. To observe the breakdown flash events, the
fibre-coupled detectors APD1 and
APD2 are optically coupled through free space by a pair of reflective collimators (RC1 and RC2) with an overall transmission of $89\%$ (including fibre losses). The reflective collimators are placed $\approx 0.2\,\rm{meters}$ apart and each one is connected to a counting module through $3\,\rm{meters}$ of optical fibre ($2\,\rm{meters}$ patchcord + $1\,\rm{meter}$ fibre in the detector module).

%The effective optical path between the two APDs is $\approx6.2\,\rm{m}$ ($6\,\rm{m}$ fibre $ + 0.2\,\rm{m}$ free space) which corresponds to a photon traveling time of $\Delta t\,\approx 31\,\rm{ns}$ between the two APDs.

The output signals from the two APDs are connected to two channels of an
oscilloscope (Lecroy Waverunner 640 Zi), which triggers when a signal is
received from APD2. Once being triggered, the oscilloscope records the arrival
time of a signal from APD1 with respect to the trigger event within the next
$250\,\rm{ns}$ with a time resolution of $100\,\rm{ps}$. An adjustable
electrical delay is applied to APD1 to offset the signal arrival time such
that only positive time differences for all interesting events are recorded by
the oscilloscope. 
%An electrical delay of $\Delta t' \approx 127\,\rm{ns}$ is applied to APD1. %the delay is later measured to be $\approx 127\,\rm{ns}$???%

The experimental setup is kept in the dark such that the breakdown flash is
only caused by dark breakdown events in the APDs. A dark breakdown event is a
thermally induced avalanche breakdown in the APD, hence it emits the same
breakdown flash light as what would be generated in a photodetection
event~\cite{Shockley1952}.  With the setup shown in
Fig.~\ref{fig:setup_collimator}(a), we observe single detector event rates of $(1.00\,\pm0.016)\times10^{4}\,\rm{s^{-1}}$ for APD1, and $(0.533\,\pm0.019)\times10^4\,\rm{s^{-1}}$ for APD2.

When there is a breakdown event in APD2 at $t=0\,\rm{s}$, the oscilloscope is
triggered. Such an event causes a breakdown flash that arrives at APD1 after a
traveling time $\Delta t$ in the optical path. This generates a signal from
APD1 which is delayed by $\Delta t'$ due to the electrical delay connected to
APD1. The signal is timestamped by the oscilloscope at $t=\Delta t + \Delta
t'$, which indicates a breakdown flash emitted from APD2 and detected by
APD1. Alternatively, a breakdown event detected in APD1 at $t =-\Delta t$
causes a breakdown flash that reaches APD2 at $t=0$ and triggers the
oscilloscope. The corresponding breakdown signal from APD1 reaches the
oscilloscope and is recorded at $t=\Delta t'-\Delta t$, indicating a breakdown flash event from APD1 detected by APD2.

\begin{figure}%[!ht]
\centering%\includegraphics[width=0.7\columnwidth]{./plots/flash_histogram/flash_histogram_boxes}
\includegraphics[width=\columnwidth]{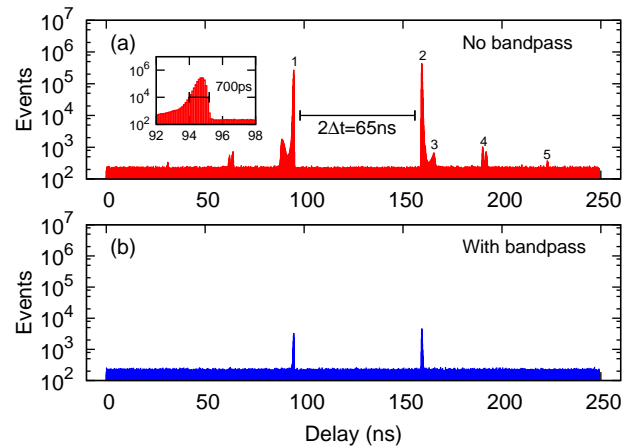}
\caption{(a) Histogram of signal arrival times from APD1 recorded by an
  oscilloscope. Peak 1 corresponds to APD1 emitting a breakdown flash that
  detected by APD2 (path A-B-C-D), peak 2 to the reverse direction (path
  D-C-B-A). Peak 3 is suspected to be due to the afterpulsing of APD1. Peaks 4
  and 5 are due to the back reflection of breakdown flash light at fibre
  joints (paths A-B-C-D-C/B-D and D-C-B-A-B/C-A). (b) Same measurement, but
  with a bandpass filter in the optical path. The number of breakdown flash events is suppressed by a factor of over 100. An integration time of $12\,\rm{hours}$ is used for both measurements.}
\label{fig:flash_histogram}
\end{figure}

Fig.~\ref{fig:flash_histogram} (a) shows the histogram of the event timings
recorded by the oscilloscope over an integration time of
$12\,\rm{hours}$. Peak 1 (located at $t_1=\Delta t' - \Delta t \approx
95\,\rm{ns}$) and peak 2 (located at $t_2=\Delta t' + \Delta t \approx 159
\,\rm{ns}$) correspond to breakdown flash events between the two APDs along
paths A-B-C-D and D-C-B-A, respectively. Each peak has a full width at half
maximum (FWHM) of $\approx 700\,\rm{ps}$. The timing separation between the
two peaks is $t_2 - t_1 =2\Delta t \approx 65\,\rm{ns}$, matching twice the
optical transit time from A to D.

Peak 3 ($t\approx166\,\rm{ns}$) is suspected to be afterpulsing signals of
APD2~\cite{Marini2016}. Peak 4 ($t\approx190\,\rm{ns}$) actually consists of
two small peaks. They are possibly due to the back reflections of photons at
the reflective collimators from a secondary breakdown flash in APD1 (triggered
by flash photons from APD2), i.e., follow a path D-C-B-A-B/C-A. The timing difference between peak 4 and peak 2 is about $31\,\rm{ns}$, which corresponds to a fibre length of about $6\,\rm{meters}$ (from point A to B/C then back to A, Fig.~\ref{fig:setup_collimator} (b)). Peak 5 ($t\approx223\,\rm{ns}$) is suspected to be a tertiary breakdown from APD2 (triggered by photons from the secondary flash in APD1), as it is about $64\,\rm{ns}$ away from peak 2 and the timing difference matches a fibre length of about $12\,\rm{meters}$ (from point A to D then back to A, Fig.~\ref{fig:setup_collimator} (b)). 

This measurement was repeated with a bandpass filter (transmission profile
shown in Fig.~\ref{fig:setup_collimator} (c)) inserted between RC1 and
RC2. The events timing histogram is shown in
Fig.~\ref{fig:flash_histogram}. The two major peaks (peak 1 and peak 2) are
suppressed by factor of about 100, while the other small peaks are no longer
observable. This indicates that spectral filtering could be used as a countermeasure to effectively reduce the breakdown flash.

The recording of timing histograms with the oscilloscope does not directly
permit the determination of absolute detection rates for breakdown flash
photons, as the histogram processing disables data taking for an
unpredictable time. We therefore employ a different method involving a
hardware coincidence stage to determine the
absolute probability of detecting a breakdown flash event. 
%To achieve this, we need to obtain the absolute rate of detected
%breakdown flash events. We use a coincidence stage and a counter to measure
%the rate of breakdown flash, and the setup is shown in
(see Fig.~\ref{fig:setup_combined} (a)). For flash events emitted by APD1 and
detected by APD2, the signal from APD1 is electrically delayed by $\Delta t'$,
matching the photon traveling time $\Delta t$. Then, an initial breakdown
signal from APD1, and the breakdown flash signal from APD2 arrive at the
coincidence stage within a coincidence window of $\sim500\,\rm{ps}$. Such a
coincidence event indicates a breakdown flash emitted from APD1 detected by
APD2, and is recorded by a hardware counter, avoiding the dead time of the
oscilloscope in data processing. The number of breakdown flash events emitted
by APD2 is measured in the same manner, except that the same electrical delay
is applied to signals from APD2.

For each configuration, we continuously record the number of coincidences for
$12\,\rm{hours}$. We find a rate of $44.4 \pm 2.2\,\rm{s^{-1}}$ from APD1 to
APD2, and $22.2 \pm 1.6\,\rm{s^{-1}}$ from APD2 to APD1. Normalized by the count rate of the emitting APDs, this yields a probability of $0.44\% \pm
0.02\%$ for APD2 detecting a breakdown flash from APD1, and a probability of
$0.42\% \pm 0.03\%$ in the reverse direction. 

In comparison, these probabilities drop to $0.0049\% \pm 0.0023\%$ and
$0.0057\% \pm 0.0033\%$, respectively, when the bandpass filter is
inserted. We estimate the rate of accidental coincidences by blocking the optical path
between the APDs, yielding a rate of $0.032\pm0.057\,\rm{s^{-1}}$, with dark
count rates of $(9.55\pm0.18)\times\,10^{3}\,\rm{s^{-1}}$ and
$(5.46\pm0.20)\times\,10^{3}\,\rm{s^{-1}}$ for APD1 and APD2,
respectively. Therefore, applying spectral filtering can effectively suppress
the rate of breakdown flash by two orders of magnitude.

\section{Spectral distribution of breakdown flash}

\begin{figure}%[!ht]
\centering%\includegraphics[width=0.8\columnwidth]{./plots/setup_spectrum/setup_combined}
\includegraphics[width=\columnwidth]{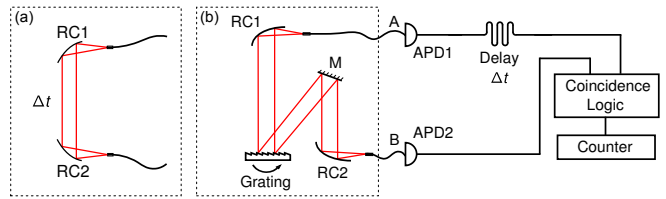}
\caption{(a) Setup for a coincidence measurement to determine the rate of detecting breakdown flashes from APD1. An electrical delay is applied to APD1 such that the dark count signal from APD1 and the breakdown flash signal from APD2 arrive at the coincidence stage at the same time. A counter is used to log the number of events per second. The setup can also measure the breakdown flash rates from APD2 with the electrical delay connected to APD2. (b) Setup for measuring the spectral distribution of the breakdown flashes. The working principle is the same as the one in (a), except that the reflective collimators are replaced by a grating monochromator to select different transmission wavelength.}
\label{fig:setup_combined}
\end{figure}

The spectral information available from the bandpass experiment is somewhat
limited. We therefore analyze the spectral distribution of the breakdown flash
light with the setup shown in Fig.~\ref{fig:setup_combined} (b). A
monochromator consisting of a reflective grating ($600\,\rm{lines/mm}$, blazed
at $1.25\,\rm{\mu m}$) and a pair of reflective collimators (RC1 and RC2) is
inserted in the optical path between the two APDs. The grating is rotated to
select the transmission wavelength between them. To estimate the spectral
resolution of the monochromator, we measure the instrument response to a
$1310\,\rm{nm}$ single mode diode laser, and find a full width at half maximum
(FWHM) of $3.3\,\rm{nm}$. For the first-order diffraction of the same
$1310\,\rm{nm}$ light, we observe a transmission of $51\%$ from point A to B in Fig.~\ref{fig:setup_combined} (b).

\begin{figure}%[!ht]
\centering%\includegraphics[width=0.68\columnwidth]{./plots/flash_spectrum/flash_spectrum}
\includegraphics[width=\columnwidth]{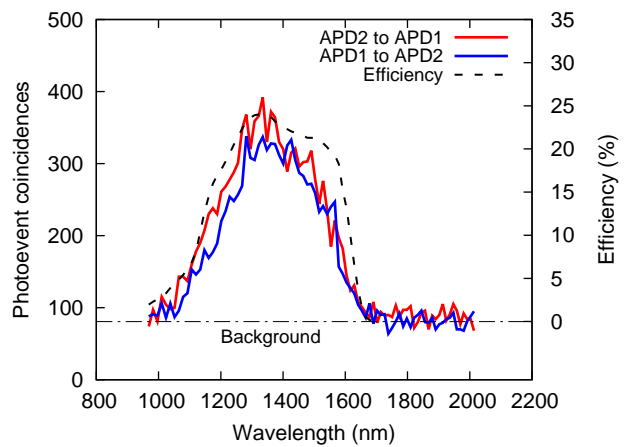}
\caption{Spectral distribution of the InGaAs APD breakdown flash. The
  integration time for each data point is $30 \,\rm{minutes}$. We record cases
  where APD1 emits a breakdown flash that is detected by APD2 and vice
  versa. The two spectra range from $1000\,\rm{nm}$ to $1600\,\rm{nm}$ and
  peak at about $1300 \,\rm{nm}$. The dashed line indicates the background
  due to accidental coincidences.}
\label{fig:flash_spectrum}
\end{figure}

We sampled 84 wavelengths ranging from $1000\,\rm{nm}$ to
$2000\,\rm{nm}$ with a grating angle incrementation of $0.28\,^\circ$. We perform same coincidence measurement as
with the single bandpass in the optical path, but with an integration time of
$30\,\rm{minutes}$. The results are shown in
Fig.~\ref{fig:flash_spectrum}. The coincidence events span a wide range from
$1000\,\rm{nm}$ to $1600\,\rm{nm}$, with a maximum at about
$1300\,\rm{nm}$. These results are not corrected for the transmission
efficiency of the monochromator, nor the wavelength-dependent detection
efficiencies of the two
APDs. However, the observed spectra (Fig.~\ref{fig:flash_spectrum}, left axis)
follow closely the spectral dependency of the nominal quantum
efficiency~\cite{datasheet} (right axis). We are not able to detect
spectral components outside the 1000\,nm-1650\,nm band. The close match of
spectral sensitivity and observed spectrum of the flash suggests that the
spectrum could be relatively flat over the whole region we are able to
observe, and could even extend beyond that sensitivity range. A more
comprehensive measurement of the actual spectrum would require more wide-band
photodetectors. The recent progress with  superconducting nanowire
detectors~\cite{Gol'tsman2001} would make these devices a good choice for such
a measurement.

%We perform a coincidence measurement to record the number of breakdown flash events at different wavelengths. The output NIM signals are adjusted to have a pulse width of $\sim 500\,\rm{ps}$ and are both sent to a coincidence stage. An electrical delay is applied to ADP1 that matches the photon travelling time $\Delta t \approx 48\,\rm{ns}$ between the two detectors. When a count is fired in APD1, it triggers a breakdown flash detected by APD2 after $\Delta t \,\rm{ns}$, the two signal pulses will arrive at the coincidence stage at the same time and creat a coincidence pulse. This is registered as a coincidence event and is recorded by a counter unit. Alternatively, if the delay is applied to APD2 instead of APD1, the setup measures number of coincidence events in which ADP1 detects breakdown flashes emitted from APD2.

%We scanned over a range from $1000\,\rm{nm}$ to $2000\,\rm{nm}$ and for each point, we count the number of coincidences over an  integration time of $30\,\rm{minutes}$. The results are shown in Fig.~\ref{fig:flash_spectrum}. We observe coincidence events over a wide range from $1000 \,\rm{nm}$ to $1600 \,\rm{nm}$. The number of events reaches its maximum at about $1300 \,\rm{nm}$. This result is uncorrected for the transmission efficiency of the grating and the detection efficiencies of the two APDs. The cut-off wavelengths of the spectra matches the detector efficiency cut-off of the APDs under testing~\cite{datasheet}, hence we suspect that the actual spectra of the breakdown flash is wider than what we observed in Fig.~\ref{fig:flash_spectrum}.

\section{Conclusion}
Commercial InGaAs single-photon counting modules do
show breakdown flash, similar to their silicon
counterparts~\cite{Kurtsiefer2001}.
We characterized the breakdown flash from two such devices using a coincidence
measurement, and obtain a lower bound for the probability of detecting a
breakdown flash of $\approx 0.4\%$. Given that these APDs have a nominal
detection 
efficiency of about $10\%$, the breakdown flash could contain at least 0.04
photons emerging from the fiber connector of the devices. This may result in a
considerable amount of information leakage that has to be considered in
practical QKD implementations.

In some sense, this should not come as a surprise, as light emission for
electron-hole recombinations in direct bandgap semiconductors like InGaAs is
more likely to happen than in indirect bandgap semiconductors like
silicon. However, a
direct comparison with photon numbers due to the breakdown flash in different
APD types based on our observation is not directly obvious: the overall
breakdown flash photon number is likely to be proportional to the charge
released in a 
breakdown, which is significantly smaller in InGaAs APD than in silicon APDs
due to the lower excess voltage above breakdown. Another unknown in the
experimental observations we have access to is the different coupling
efficiencies of the diodes in this experiment (diodes are connected to
multi-mode fibers with a small core diameter) and free space diodes used in
earlier experiments~\cite{Kurtsiefer2001}.

The spectral distribution of the breakdown flash appears to be
relatively wide. Thus, a spectral filter in front of an APD is a suitable
countermeasure to prevent potential information leakage through the breakdown
flash in a quantum key distribution scenario. In that narrow spectral window,
the detection efficiency can be assumed to 
be constant. With a conservative estimation of the detector efficiency, the
observed detected breakdown flash probability can then be used to provide an
upper bound for estimating the number of photons being transmitted to the
optical channel due the breakdown flash.\vspace*{5mm}

\section*{Acknowledgements}
This research is supported by the Ministry of Education, Singapore, and the
National Research Foundation, Prime Minister's Office, Singapore, partly under its Corporate Laboratory@University Scheme, National University of Singapore, and Singapore Telecommunications Ltd.

\bibliographystyle{apsrev4-1}
%\bibliography{reference.bib}

\begin{thebibliography}{17}%
\makeatletter
\providecommand \@ifxundefined [1]{%
 \@ifx{#1\undefined}
}%
\providecommand \@ifnum [1]{%
 \ifnum #1\expandafter \@firstoftwo
 \else \expandafter \@secondoftwo
 \fi
}%
\providecommand \@ifx [1]{%
 \ifx #1\expandafter \@firstoftwo
 \else \expandafter \@secondoftwo
 \fi
}%
\providecommand \natexlab [1]{#1}%
\providecommand \enquote  [1]{``#1''}%
\providecommand \bibnamefont  [1]{#1}%
\providecommand \bibfnamefont [1]{#1}%
\providecommand \citenamefont [1]{#1}%
\providecommand \href@noop [0]{\@secondoftwo}%
\providecommand \href [0]{\begingroup \@sanitize@url \@href}%
\providecommand \@href[1]{\@@startlink{#1}\@@href}%
\providecommand \@@href[1]{\endgroup#1\@@endlink}%
\providecommand \@sanitize@url [0]{\catcode `\\12\catcode `\$12\catcode
  `\&12\catcode `\#12\catcode `\^12\catcode `\_12\catcode `\%12\relax}%
\providecommand \@@startlink[1]{}%
\providecommand \@@endlink[0]{}%
\providecommand \url  [0]{\begingroup\@sanitize@url \@url }%
\providecommand \@url [1]{\endgroup\@href {#1}{\urlprefix }}%
\providecommand \urlprefix  [0]{URL }%
\providecommand \Eprint [0]{\href }%
\providecommand \doibase [0]{http://dx.doi.org/}%
\providecommand \selectlanguage [0]{\@gobble}%
\providecommand \bibinfo  [0]{\@secondoftwo}%
\providecommand \bibfield  [0]{\@secondoftwo}%
\providecommand \translation [1]{[#1]}%
\providecommand \BibitemOpen [0]{}%
\providecommand \bibitemStop [0]{}%
\providecommand \bibitemNoStop [0]{.\EOS\space}%
\providecommand \EOS [0]{\spacefactor3000\relax}%
\providecommand \BibitemShut  [1]{\csname bibitem#1\endcsname}%
\let\auto@bib@innerbib\@empty
%</preamble>
\bibitem [{\citenamefont {Bennett}\ and\ \citenamefont
  {Brassard}(1984)}]{Bennett1984}%
  \BibitemOpen
  \bibfield  {author} {\bibinfo {author} {\bibfnamefont {C.~H.}\ \bibnamefont
  {Bennett}}\ and\ \bibinfo {author} {\bibfnamefont {G.}~\bibnamefont
  {Brassard}},\ }in\ \href@noop {} {\emph {\bibinfo {booktitle} {International
  Conference on Computers, Systems \& Signal Processing, Bangalore, India, Dec
  9-12, 1984}}}\ (\bibinfo {year} {1984})\ pp.\ \bibinfo {pages}
  {175--179}\BibitemShut {NoStop}%
\bibitem [{\citenamefont {Buttler}\ \emph {et~al.}(1998)\citenamefont
  {Buttler}, \citenamefont {Hughes}, \citenamefont {Kwiat}, \citenamefont
  {Lamoreaux}, \citenamefont {Luther}, \citenamefont {Morgan}, \citenamefont
  {Nordholt}, \citenamefont {Peterson},\ and\ \citenamefont
  {Simmons}}]{Buttler1998}%
  \BibitemOpen
  \bibfield  {author} {\bibinfo {author} {\bibfnamefont {W.~T.}\ \bibnamefont
  {Buttler}}, \bibinfo {author} {\bibfnamefont {R.~J.}\ \bibnamefont {Hughes}},
  \bibinfo {author} {\bibfnamefont {P.~G.}\ \bibnamefont {Kwiat}}, \bibinfo
  {author} {\bibfnamefont {S.~K.}\ \bibnamefont {Lamoreaux}}, \bibinfo {author}
  {\bibfnamefont {G.~G.}\ \bibnamefont {Luther}}, \bibinfo {author}
  {\bibfnamefont {G.~L.}\ \bibnamefont {Morgan}}, \bibinfo {author}
  {\bibfnamefont {J.~E.}\ \bibnamefont {Nordholt}}, \bibinfo {author}
  {\bibfnamefont {C.~G.}\ \bibnamefont {Peterson}}, \ and\ \bibinfo {author}
  {\bibfnamefont {C.~M.}\ \bibnamefont {Simmons}},\ }\href {\doibase
  10.1103/PhysRevLett.81.3283} {\bibfield  {journal} {\bibinfo  {journal}
  {Phys. Rev. Lett.}\ }\textbf {\bibinfo {volume} {81}},\ \bibinfo {pages}
  {3283} (\bibinfo {year} {1998})}\BibitemShut {NoStop}%
\bibitem [{\citenamefont {Marcikic}\ \emph {et~al.}(2006)\citenamefont
  {Marcikic}, \citenamefont {Lamas-Linares},\ and\ \citenamefont
  {Kurtsiefer}}]{Marcikic2006}%
  \BibitemOpen
  \bibfield  {author} {\bibinfo {author} {\bibfnamefont {I.}~\bibnamefont
  {Marcikic}}, \bibinfo {author} {\bibfnamefont {A.}~\bibnamefont
  {Lamas-Linares}}, \ and\ \bibinfo {author} {\bibfnamefont {C.}~\bibnamefont
  {Kurtsiefer}},\ }\href@noop {} {\bibfield  {journal} {\bibinfo  {journal}
  {Appl. Phys. Lett.}\ }\textbf {\bibinfo {volume} {89}},\ \bibinfo {pages}
  {101122} (\bibinfo {year} {2006})} \BibitemShut
  {NoStop}%
\bibitem [{\citenamefont {Jouguet}\ \emph {et~al.}(2012)\citenamefont
  {Jouguet}, \citenamefont {Kunz-Jacques}, \citenamefont {Leverrier},
  \citenamefont {Grangier},\ and\ \citenamefont {Diamanti}}]{Jouguet2012}%
  \BibitemOpen
  \bibfield  {author} {\bibinfo {author} {\bibfnamefont {P.}~\bibnamefont
  {Jouguet}}, \bibinfo {author} {\bibfnamefont {S.}~\bibnamefont
  {Kunz-Jacques}}, \bibinfo {author} {\bibfnamefont {A.}~\bibnamefont
  {Leverrier}}, \bibinfo {author} {\bibfnamefont {P.}~\bibnamefont {Grangier}},
  \ and\ \bibinfo {author} {\bibfnamefont {E.}~\bibnamefont {Diamanti}},\
  }\href@noop {} {\bibfield  {journal} {\bibinfo  {journal} {Nature Photonics}\
  }\textbf {\bibinfo {volume} {7}},\ \bibinfo {pages} {378} (\bibinfo {year}
  {2012})} \BibitemShut
  {NoStop}%
\bibitem [{\citenamefont {Townsend}(1999)}]{Townsend1999}%
  \BibitemOpen
  \bibfield  {author} {\bibinfo {author} {\bibfnamefont {P.~D.}\ \bibnamefont
  {Townsend}},\ }in\ \href {\doibase 10.1109/OFC.1999.766020} {\emph {\bibinfo
  {booktitle} {OFC/IOOC . Technical Digest. Optical Fiber Communication
  Conference, 1999, and the International Conference on Integrated Optics and
  Optical Fiber Communication}}},\ Vol.~\bibinfo {volume} {4}\ (\bibinfo {year}
  {1999})\ pp.\ \bibinfo {pages} {141--143}\BibitemShut {NoStop}%
\bibitem [{\citenamefont {Bourennane}\ \emph {et~al.}(1999)\citenamefont
  {Bourennane}, \citenamefont {Gibson}, \citenamefont {Karlsson}, \citenamefont
  {Hening}, \citenamefont {Jonsson}, \citenamefont {Tsegaye}, \citenamefont
  {Ljunggren},\ and\ \citenamefont {Sundberg}}]{Bourennane1999}%
  \BibitemOpen
  \bibfield  {author} {\bibinfo {author} {\bibfnamefont {M.}~\bibnamefont
  {Bourennane}}, \bibinfo {author} {\bibfnamefont {F.}~\bibnamefont {Gibson}},
  \bibinfo {author} {\bibfnamefont {A.}~\bibnamefont {Karlsson}}, \bibinfo
  {author} {\bibfnamefont {A.}~\bibnamefont {Hening}}, \bibinfo {author}
  {\bibfnamefont {P.}~\bibnamefont {Jonsson}}, \bibinfo {author} {\bibfnamefont
  {T.}~\bibnamefont {Tsegaye}}, \bibinfo {author} {\bibfnamefont
  {D.}~\bibnamefont {Ljunggren}}, \ and\ \bibinfo {author} {\bibfnamefont
  {E.}~\bibnamefont {Sundberg}},\ }\href {\doibase 10.1364/OE.4.000383}
  {\bibfield  {journal} {\bibinfo  {journal} {Opt. Express}\ }\textbf {\bibinfo
  {volume} {4}},\ \bibinfo {pages} {383} (\bibinfo {year} {1999})}\BibitemShut
  {NoStop}%
\bibitem [{\citenamefont {Ribordy}\ \emph {et~al.}(2000)\citenamefont
  {Ribordy}, \citenamefont {Gautier}, \citenamefont {Gisin}, \citenamefont
  {Guinnard},\ and\ \citenamefont {Zbinden}}]{Ribordy2000}%
  \BibitemOpen
  \bibfield  {author} {\bibinfo {author} {\bibfnamefont {G.}~\bibnamefont
  {Ribordy}}, \bibinfo {author} {\bibfnamefont {J.~D.}\ \bibnamefont
  {Gautier}}, \bibinfo {author} {\bibfnamefont {N.}~\bibnamefont {Gisin}},
  \bibinfo {author} {\bibfnamefont {O.}~\bibnamefont {Guinnard}}, \ and\
  \bibinfo {author} {\bibfnamefont {H.}~\bibnamefont {Zbinden}},\ }\href@noop
  {} {\bibfield  {journal} {\bibinfo  {journal} {Journal of Modern Optics}\
  }\textbf {\bibinfo {volume} {47}},\ \bibinfo {pages} {517} (\bibinfo {year}
  {2000})}\BibitemShut {NoStop}%
\bibitem [{\citenamefont {Hughes}\ \emph {et~al.}(2000)\citenamefont {Hughes},
  \citenamefont {Morgan},\ and\ \citenamefont {Peterson}}]{Hughes2000}%
  \BibitemOpen
  \bibfield  {author} {\bibinfo {author} {\bibfnamefont {R.~J.}\ \bibnamefont
  {Hughes}}, \bibinfo {author} {\bibfnamefont {G.~L.}\ \bibnamefont {Morgan}},
  \ and\ \bibinfo {author} {\bibfnamefont {C.~G.}\ \bibnamefont {Peterson}},\
  }\href {\doibase 10.1080/09500340008244058} {\bibfield  {journal} {\bibinfo
  {journal} {Journal of Modern Optics}\ }\textbf {\bibinfo {volume} {47}},\
  \bibinfo {pages} {533} (\bibinfo {year} {2000})} \BibitemShut {NoStop}%
\bibitem [{\citenamefont {Muller}\ \emph {et~al.}(1997)\citenamefont {Muller},
  \citenamefont {Herzog}, \citenamefont {Huttner}, \citenamefont {Tittel},
  \citenamefont {Zbinden},\ and\ \citenamefont {Gisin}}]{Muller1997}%
  \BibitemOpen
  \bibfield  {author} {\bibinfo {author} {\bibfnamefont {A.}~\bibnamefont
  {Muller}}, \bibinfo {author} {\bibfnamefont {T.}~\bibnamefont {Herzog}},
  \bibinfo {author} {\bibfnamefont {B.}~\bibnamefont {Huttner}}, \bibinfo
  {author} {\bibfnamefont {W.}~\bibnamefont {Tittel}}, \bibinfo {author}
  {\bibfnamefont {H.}~\bibnamefont {Zbinden}}, \ and\ \bibinfo {author}
  {\bibfnamefont {N.}~\bibnamefont {Gisin}},\ }\href@noop {} {\bibfield
  {journal} {\bibinfo  {journal} {Applied Physics Letters}\ }\textbf {\bibinfo
  {volume} {70}},\ \bibinfo {pages} {793} (\bibinfo {year} {1997})}\BibitemShut
  {NoStop}%
\bibitem [{\citenamefont {Stucki}\ \emph {et~al.}(2001)\citenamefont {Stucki},
  \citenamefont {Ribordy}, \citenamefont {Stefanov}, \citenamefont {Zbinden},
  \citenamefont {Rarity},\ and\ \citenamefont {Wall}}]{Stucki2001}%
  \BibitemOpen
  \bibfield  {author} {\bibinfo {author} {\bibfnamefont {D.}~\bibnamefont
  {Stucki}}, \bibinfo {author} {\bibfnamefont {G.}~\bibnamefont {Ribordy}},
  \bibinfo {author} {\bibfnamefont {A.}~\bibnamefont {Stefanov}}, \bibinfo
  {author} {\bibfnamefont {H.}~\bibnamefont {Zbinden}}, \bibinfo {author}
  {\bibfnamefont {J.~G.}\ \bibnamefont {Rarity}}, \ and\ \bibinfo {author}
  {\bibfnamefont {T.}~\bibnamefont {Wall}},\ }\href {\doibase
  10.1080/09500340108240900} {\bibfield  {journal} {\bibinfo  {journal}
  {Journal of Modern Optics}\ }\textbf {\bibinfo {volume} {48}},\ \bibinfo
  {pages} {1967} (\bibinfo {year} {2001})} \BibitemShut {NoStop}%
\bibitem [{\citenamefont {Hadfield}(2009)}]{Hadfield2009}%
  \BibitemOpen
  \bibfield  {author} {\bibinfo {author} {\bibfnamefont {R.}~\bibnamefont
  {Hadfield}},\ }\href {\doibase 10.1038/nphoton.2009.230} {\bibfield
  {journal} {\bibinfo  {journal} {Nature Photonics}\ }\textbf {\bibinfo
  {volume} {3}},\ \bibinfo {pages} {696} (\bibinfo {year} {2009})}\BibitemShut
  {NoStop}%
\bibitem [{\citenamefont {Zhang}\ \emph {et~al.}(2015)\citenamefont {Zhang},
  \citenamefont {Itzler}, \citenamefont {Zbinden},\ and\ \citenamefont
  {Pan}}]{Zhang2015}%
  \BibitemOpen
  \bibfield  {author} {\bibinfo {author} {\bibfnamefont {J.}~\bibnamefont
  {Zhang}}, \bibinfo {author} {\bibfnamefont {M.~A.}\ \bibnamefont {Itzler}},
  \bibinfo {author} {\bibfnamefont {H.}~\bibnamefont {Zbinden}}, \ and\
  \bibinfo {author} {\bibfnamefont {J.-W.}\ \bibnamefont {Pan}},\ }\href@noop
  {} {\bibfield  {journal} {\bibinfo  {journal} {Light: Science \&
  Applications}\ }\textbf {\bibinfo {volume} {4}},\ \bibinfo {pages} {e286}
  (\bibinfo {year} {2015})} \BibitemShut {NoStop}%
\bibitem [{\citenamefont {Marini}\ \emph {et~al.}(2016)\citenamefont {Marini},
  \citenamefont {Camphausen}, \citenamefont {Xiong}, \citenamefont {Eggleton},\
  and\ \citenamefont {Palomba}}]{Marini2016}%
  \BibitemOpen
  \bibfield  {author} {\bibinfo {author} {\bibfnamefont {L.}~\bibnamefont
  {Marini}}, \bibinfo {author} {\bibfnamefont {R.}~\bibnamefont {Camphausen}},
  \bibinfo {author} {\bibfnamefont {C.}~\bibnamefont {Xiong}}, \bibinfo
  {author} {\bibfnamefont {B.}~\bibnamefont {Eggleton}}, \ and\ \bibinfo
  {author} {\bibfnamefont {S.}~\bibnamefont {Palomba}},\ }\href {\doibase
  10.1364/acoft.2016.aw5c.4} {\bibfield  {journal} {\bibinfo  {journal}
  {Photonics and Fiber Technology 2016 (ACOFT, BGPP, NP)}\ } (\bibinfo {year}
  {2016})}\BibitemShut {NoStop}%
\bibitem [{\citenamefont {Kurtsiefer}\ \emph {et~al.}(2001)\citenamefont
  {Kurtsiefer}, \citenamefont {Zarda}, \citenamefont {Mayer},\ and\
  \citenamefont {Weinfurter}}]{Kurtsiefer2001}%
  \BibitemOpen
  \bibfield  {author} {\bibinfo {author} {\bibfnamefont {C.}~\bibnamefont
  {Kurtsiefer}}, \bibinfo {author} {\bibfnamefont {P.}~\bibnamefont {Zarda}},
  \bibinfo {author} {\bibfnamefont {S.}~\bibnamefont {Mayer}}, \ and\ \bibinfo
  {author} {\bibfnamefont {H.}~\bibnamefont {Weinfurter}},\ }\href {\doibase
  10.1080/09500340108240905} {\bibfield  {journal} {\bibinfo  {journal}
  {Journal of Modern Optics}\ }\textbf {\bibinfo {volume} {48}},\ \bibinfo
  {pages} {2039} (\bibinfo {year} {2001})}\BibitemShut {NoStop}%
\bibitem [{\citenamefont {Shockley}\ and\ \citenamefont
  {Read}(1952)}]{Shockley1952}%
  \BibitemOpen
  \bibfield  {author} {\bibinfo {author} {\bibfnamefont {W.}~\bibnamefont
  {Shockley}}\ and\ \bibinfo {author} {\bibfnamefont {W.~T.}\ \bibnamefont
  {Read}},\ }\href {\doibase 10.1103/PhysRev.87.835} {\bibfield  {journal}
  {\bibinfo  {journal} {Phys. Rev.}\ }\textbf {\bibinfo {volume} {87}},\
  \bibinfo {pages} {835} (\bibinfo {year} {1952})}\BibitemShut {NoStop}%
\bibitem [{dat()}]{datasheet}%
  \BibitemOpen
  \href@noop {} {\emph {\bibinfo {title} {Data sheet for ID220 Infrared
  Single-Photon Detector}}},\ \bibinfo {organization} {ID
  Quantique}\BibitemShut {NoStop}%
\bibitem [{\citenamefont {Gol'tsman}\ \emph {et~al.}(2001)\citenamefont
  {Gol'tsman}, \citenamefont {Okunev}, \citenamefont {Chulkova}, \citenamefont
  {Lipatov}, \citenamefont {Semenov}, \citenamefont {Smirnov}, \citenamefont
  {Voronov}, \citenamefont {Dzardanov}, \citenamefont {Williams},\ and\
  \citenamefont {Sobolewski}}]{Gol'tsman2001}%
  \BibitemOpen
  \bibfield  {author} {\bibinfo {author} {\bibfnamefont {G.~N.}\ \bibnamefont
  {Gol'tsman}}, \bibinfo {author} {\bibfnamefont {O.}~\bibnamefont {Okunev}},
  \bibinfo {author} {\bibfnamefont {G.}~\bibnamefont {Chulkova}}, \bibinfo
  {author} {\bibfnamefont {A.}~\bibnamefont {Lipatov}}, \bibinfo {author}
  {\bibfnamefont {A.}~\bibnamefont {Semenov}}, \bibinfo {author} {\bibfnamefont
  {K.}~\bibnamefont {Smirnov}}, \bibinfo {author} {\bibfnamefont
  {B.}~\bibnamefont {Voronov}}, \bibinfo {author} {\bibfnamefont
  {A.}~\bibnamefont {Dzardanov}}, \bibinfo {author} {\bibfnamefont
  {C.}~\bibnamefont {Williams}}, \ and\ \bibinfo {author} {\bibfnamefont
  {R.}~\bibnamefont {Sobolewski}},\ }\href {\doibase 10.1063/1.1388868}
  {\bibfield  {journal} {\bibinfo  {journal} {Applied Physics Letters}\
  }\textbf {\bibinfo {volume} {79}},\ \bibinfo {pages} {705} (\bibinfo {year}
  {2001})} \BibitemShut {NoStop}%
\end{thebibliography}

\end{document}